# A FULLY-PARAMETERIZED FEM MODEL FOR ELECTROMAGNETIC OPTIMIZATION OF AN RF-MEMS WAFER-LEVEL PACKAGE

J. Iannacci[1,2], J. Tian[2], R. Gaddi[1], A. Gnudi[1], and M. Bartek[2]

1) ARCES-DEIS Università di Bologna, Via Risorgimento 2, 40123 Bologna, Italy

2) HiTeC-DIMES, Delft University of Technology, Mekelweg 4, 2628 CD Delft, the Netherlands

E-mail: jiannacci@deis.unibo.it

Phone: +39 (0)51 20 93049    Fax: +39 (0)51 20 93779

## ABSTRACT

In this paper we present our efforts in characterizing and optimizing the influence of a Wafer-Level Packaging (WLP) solution on the electromagnetic behaviour of RF-MEMS devices. To this purpose, a fully parameterized FEM model of a packaged Coplanar Waveguide (CPW) is presented in order to optimize all the technology degrees of freedom (DoF's) made available by the fabrication process of the capping part. The model is implemented within the Ansoft HFSS[TM] electromagnetic simulator, after its validation against experimental data. Moreover, a simulation approach of a capped RF-MEMS varactor is shown. It is implemented in the Spectre[©] simulator within Cadence[©] environment. The MEMS part is treated by means of a compact model library implemented in VerilogA[©] language. A lumped elements network accounting for the parasitics surrounding the intrinsic RF-MEMS varactor is extracted from experimental data. Finally, the S-parameters description of the package, obtained by Ansoft HFSS[TM] simulations, is included in the Spectre[©] schematic.

## 1. INTRODUCTION

Packaging has recently been identified as the enabling factor of electronic system performance enhancement and consequently, its technology has gained considerable attention [1]. Concerning MEMS devices, the packaging plays even a more critical role. Indeed, since MEMS devices contain movable parts, like very thin suspended membranes, they need appropriate protection. Factors like shock, moisture and dust particles can partially or totally compromise the proper functionality of such devices. Moreover, when dealing with MEMS for Radio Frequency (RF) applications, additional issues related to the packaging come in [2]. For encapsulation of MEMS structures a protective substrate is usually employed (Wafer-Level Packaging). This additional part introduces parasitic (capacitive and inductive) effects related, for example, to vertical through-wafer vias. In addition, after the wafer-to-wafer bonding, the capping part is very close to the device substrate. Hence, the reduced air gap causes electromagnetic couplings between devices that introduce additional losses and mismatch. Parasitics introduced by the application of the capping part have to be reduced as much as possible in order not to compromise the RF functionality of the packaged MEMS devices.

## 2. ANSOFT HFSS[TM] VALIDATION

Before exploiting Ansoft HFSS[TM] to optimize the package design with respect to its electromagnetic behavior, this has first to be validated against experimental data. To this purpose, simulated results of capped 50 Ω CPW's and shorts have been compared with measurements. The package fabrication is based on the etching of through-wafer vias subsequently filled with Copper and is provided by the DIMES Technology Centre [3].

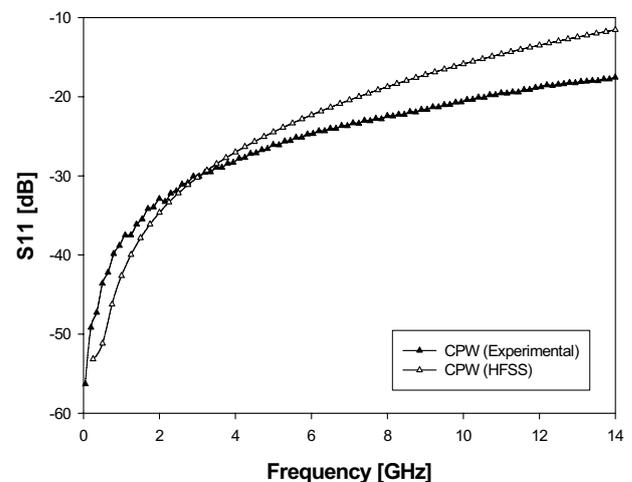

Fig. 1: Comparison of the experimental and simulated S11 parameter for a capped CPW.

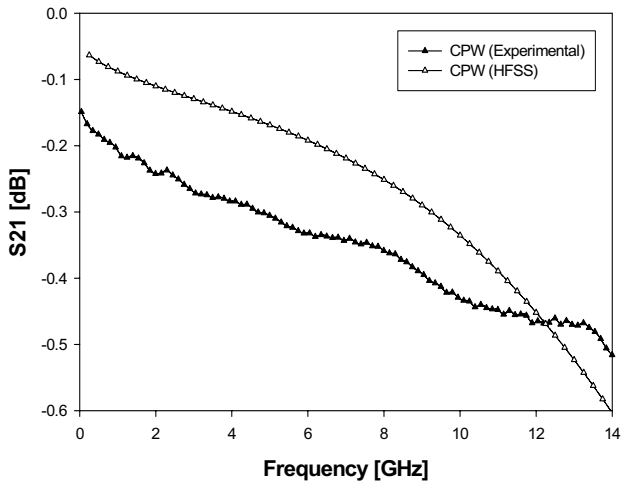

Fig. 2: Comparison of the experimental and simulated S21 parameter for a capped CPW.

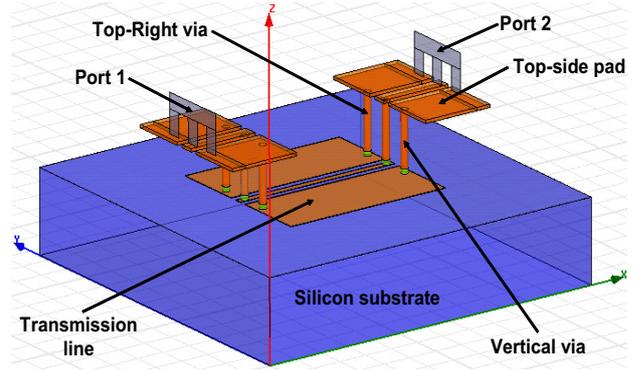

Fig. 3: Ansoft HFSS[TM] schematic view of a capped CPW. The capping substrate is not shown to get a plain view of the underneath device and of the through-wafer interconnects.

In Figures 1 and 2 the measured and simulated S11 and S21 scattering parameters for a capped CPW are compared, showing good agreement. For instance, the offset between the two curves for the reflection parameter (S11) is 5.3 dB at 12 GHz (Figure 1). On the other hand, the offset between the simulated and measured S21 parameter is 0.14 dB at 6 GHz. The CPW is 1350 µm long. Signal and ground lines width is 116 µm and 300 µm respectively and the gap is 65 µm. Package thickness is around 280 µm and through-vertical vias diameter is 50 µm. S-parameters for other packaged CPW's and shorts topologies are similar to the ones shown in previous plots. This confirms Ansoft HFSS[TM] to be a suitable tool for the accurate prediction of capped structures RF behaviour.

### 3. CAPPED LINE PARAMETERIZED MODEL

When dealing with several technology degrees of freedom (DoF's), their parameterization allows for fully automated optimization. In this work, we focus on the parasitics reduction of the package applied to 50 Ω CPW's and shorts instead of actual RF-MEMS devices. This choice is done mainly because the influence of the capping substrate is easily interpretable when applied to structures with a very simple frequency response. Indeed, in this preliminary stage of the packaging process development it would not be useful to focus directly on the cap influence on RF-performances of complete MEMS devices without knowing the general trend of each DoF within reasonable ranges.

In the HFSS[TM] parameterized model, suitable independent variables have been defined in order to describe all the technology DoF's (e.g. via diameter, capping substrate thickness, recess depth etc.). The HFSS[TM] 3D-view of a capped CPW is shown in Figure 3. In order to explain the approach the Top-Right via (Figure 3) is considered here. The relations between independent and dependent variables in the definition of the Top-Right via centre coordinates (x, y, z) are (see Figures 4 and 5):

$$x = \frac{1}{2} x_{box} + \frac{1}{2} x_{line} - x_{offset} \quad (1a)$$

$$y = \frac{1}{2} y_{box} + y_{offset} \quad (1b)$$

$$z = z_{Si} + z_{Ox} + z_{line} + z_{bump} \quad (1c)$$

where $x_{box}$ and $y_{box}$ are the substrate dimensions, $x_{offset}$ and $y_{offset}$ the centre via distance from the edge of the line and the distance between the signal via and the ground vias, respectively (Figure 4), $x_{line}$ is the line length, and $z_{Si}$, $z_{Ox}$, $z_{line}$, $z_{bump}$ are the thicknesses of the Silicon substrate, oxide layer, CPW and bumps respectively (Figure 5). As proof of concept the parameterized model is exploited to investigate the influence of two geometrical DoF's on the behaviour of the S-parameters of a CPW with a length of 1500 µm, signal and ground lines widths of 100 µm and 300 µm, respectively, and gap between the signal and ground lines of 50 µm.

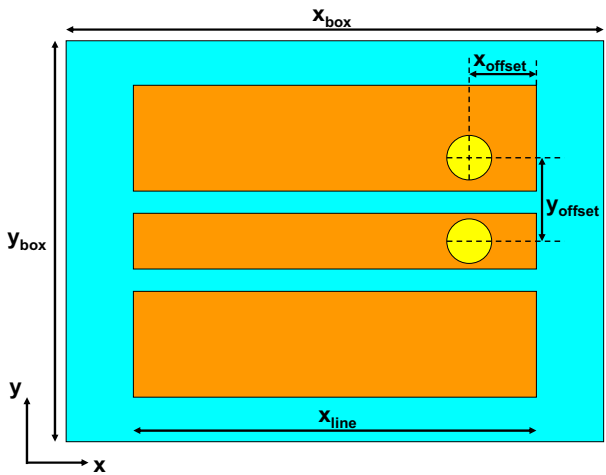

Fig. 4: Definition of the Top-Right via centre (shown in Figure 3) coordinates on the xy-plane.

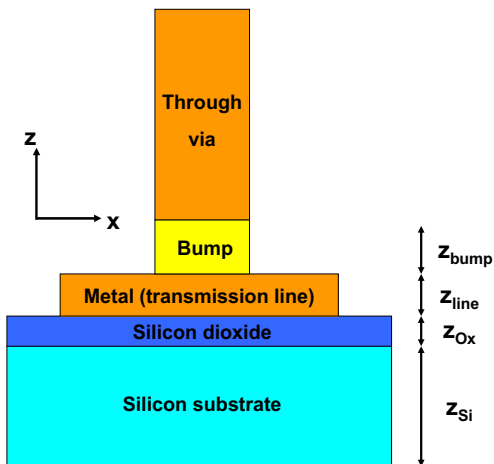

Fig. 5: Definition of the Top-Right via centre (shown in Figure 3) coordinates along the z-direction.

The two analyzed DoF's are via diameter and the y-axis distance between signal and ground vias ($y_{offset}$) shown in Figure 4. Via diameter ranges between 5 μm and 95 μm. The latter value was considered the largest achievable since the signal line is 100 μm wide. Furthermore, the y-axis distance between vias is varied from 150 μm up to 350 μm. This means that the ground vias position varies from one edge of the ground line to the opposite one. The signal pads on the package top-side are chosen with the same width as the capped ground lines. Finally, the frequency is fixed to 5 GHz. The optimization results are shown in Figure 6. On the xy-plane of the 3D plot via diameter and lateral via distance are reported while on the z-axis the transmission parameter (S21) at 5 GHz is shown. The goal of the optimization is to maximize the S21 value. By observing the plot it is noticed that the lowest values of the transmission parameter correspond to the narrowest via diameter.

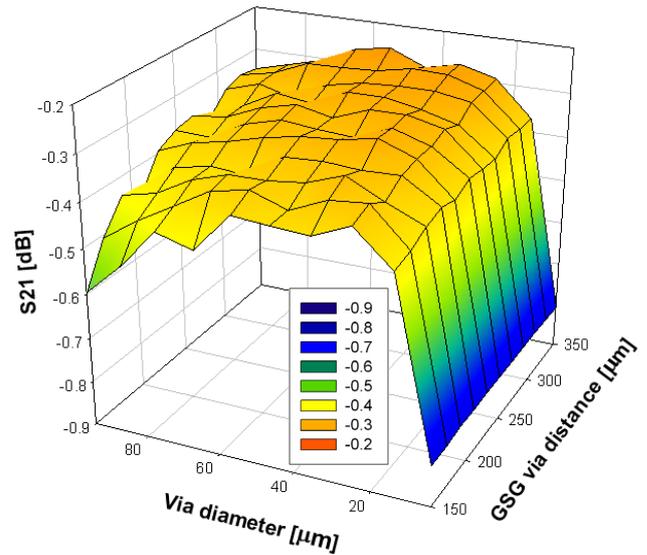

Fig. 6: 3D plot of via diameter and lateral distance optimization for a capped transmission line at 5 GHz.

Running extensive simulations, including all the technology DoF's, the ones that exhibit larger influence on the RF behaviour of various capped CPW geometries are identified. These are listed in Figure 7 together with the qualitative effect of each of them on the additional losses and mismatch. For instance, an increases of the recess depth as well as of the via diameter leads to a reduction of parasitics associated to the package, while increasing the cap height leads to an undesired increase of losses and mismatch. This allows defining appropriate ranges for the package DoF's values to ensure small parasitics. Of course, in defining these guidelines for the optimum design of the capping part, issues related to the technology are also accounted for. For example, the package thinning leads to better RF performance, but a trade-off raises with the mechanical strength of the cap itself. Accounting for this issue, the optimum package height is between 250 μm and 300 μm. As far as the substrate resistivity of the cap is concerned, various possibilities have been considered also accounting for the wafers available for the fabrication (15 Ωcm and 1, 2, 3, 4 KΩcm). From Figure 7, it appears that the use of a High-Resistivity Substrate (HRS) is to be preferred. The choice of the proper HRS value brings-up a possible trade-off between performances and costs. In this respect, since the benefits achieved with very HRS's (3−4 KΩcm) are not large with respect to the 1 KΩcm and 2 KΩcm wafers, the best choice would probably fall between one of the latter two, the use of the 2 KΩcm being the most reasonable choice.

| Capping Silicon substrate resistivity | + |
|---|---|
| Capping Silicon substrate height | - |
| Recess depth | + |
| Via diameter | + |
| GSG vias lateral distance | + |
| Bumps height | + |

Fig. 7: Summary of the most important technology DoF's in package fabrication concerning their influence on RF performance of MEMS. The signs '+' or '-' indicate whether the corresponding DoF must be increased or decreased in order to reduce the package related parasitics.

Recess height should be large but not too much according to the mechanical strength of the cap. When a 250-300 µm thick package is employed, reasonable recess depth should be around 100 µm. Finally, via diameter should be as large as possible (not less than 60-70 µm) to reduce the resistance. Also horizontal spacing between signal and ground vias has to be large (more than 250 µm) to reduce capacitive couplings. Bumps height must be large (not less than 20-40 µm) in order to keep the largest possible gap between device and capping wafers. Finally, non-critical DoF's concerning the package-related parasitics are also identified, like the oxide layer thickness on vias sidewalls, which can be chosen without particular issues. All these considerations are based on the Ansoft HFSS$^{TM}$ results of parametric simulations up to 16 GHz. All the mentioned results have been taken into account in designing an optimized package substrate for RF-MEMS devices which is currently being fabricated.

### 4. SIMULATION OF A CAPPED RF-MEMS VARACTOR

As proof of concept a Spectre$^©$ simulation of a packaged RF-MEMS varactor realized in ITC-irst technology [4] is shown in this section. The varactor is based on a central rigid plate with four suspending meander structures anchored at its corners. Meanders are preferred to simple straight beams as they allow both reducing the pull-in voltage and alleviating the effect of material residual stress [5]. The intrinsic RF-MEMS varactor electromechanical model is defined in Spectre$^©$ by assembling together elementary components (i.e. flexible beams and rigid plates) from a model library implemented in VerilogA$^©$ language [6]. A 3D-view of the fabricated MEMS varactor obtained by an optical profilometer (Veeco$^{TM}$ WYKO NT1100 DMEMS system) is shown in Figure 8. Cadence Virtuoso$^©$ Schematic of the same structure is reported in Figure 9.

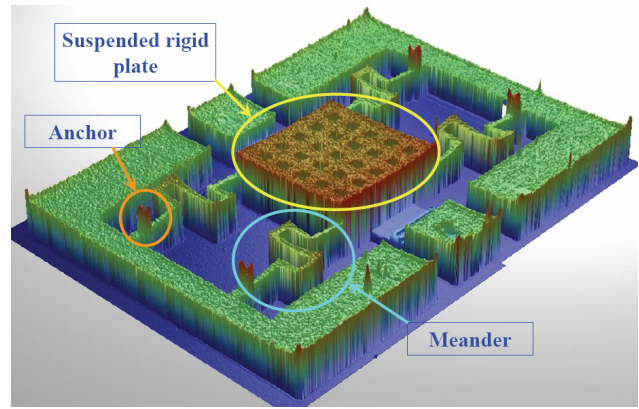

Fig. 8: 3D view of the RF-MEMS varactor fabricated in ITC-irst technology. The four meander structures connected to the central suspended plate are visible. (Veeco™ WYKO NT1100).

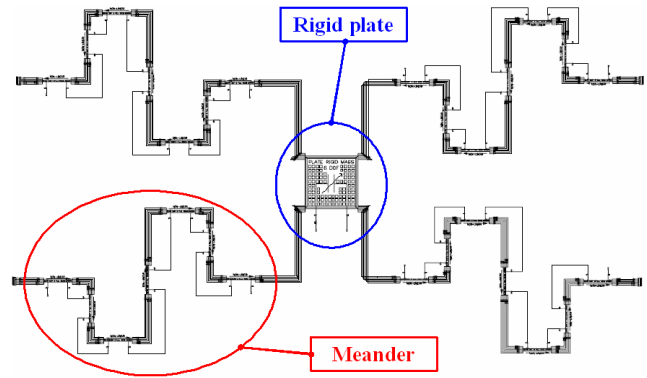

Fig. 9: Cadence Virtuoso$^©$ Schematic of the RF-MEMS varactor based on a central rigid plate suspended with four meander structures.

Moreover, a lumped elements network accounting for the parasitics surrounding the intrinsic RF-MEMS varactor (e.g. series inductance and resistance of the input and output lines) is extracted from the collected measured data following a well-known approach in microwave transistor modeling [7]. The complete lumped elements network is shown in Figure 10, where the intrinsic RF-MEMS varactor of Figure 9 is instantiated with a corresponding symbol. As the losses of the varactor are not included in the MEMS compact model in VerilogA$^©$, a conductance is connected in parallel to the intrinsic device and set to the mean measured value.

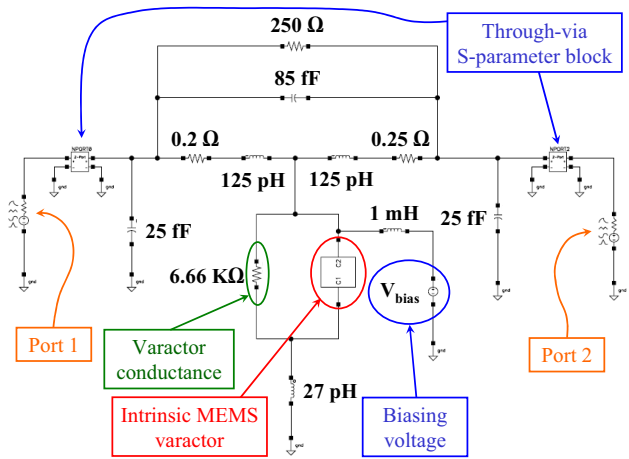

Fig. 10: Lumped elements network accounting for the parasitics surrounding the intrinsic RF-MEMS varactor. The electromagnetic influence of the package is counted in by the through-wafer vias S-parameters blocks extracted from Ansoft HFSS[TM] simulations.

The electromagnetic influence of the package is accounted for by including in the network two S-parameters blocks extracted from simulations. The structure implemented in HFSS[TM] is a set of three vertical GSG vias through a 2 KΩcm substrate. Vias are contacted both at the top and bottom ends by short CPW's leading to the two ports (see Figure 11). This is done in order not to place the ports directly on vias openings, which would neglect the discontinuities between the CPW's and vertical vias. Via diameter is 70 μm while the capping substrate (not shown in Figure 11) is 350 μm thick. The S-parameters are exported in Touchstone[©] format and linked in Spectre[©] to the two blocks shown in Figure 10. HFSS[TM] simulation is performed up to 10 GHz as the lumped network accounting for the parasitics has not been extracted beyond this frequency.

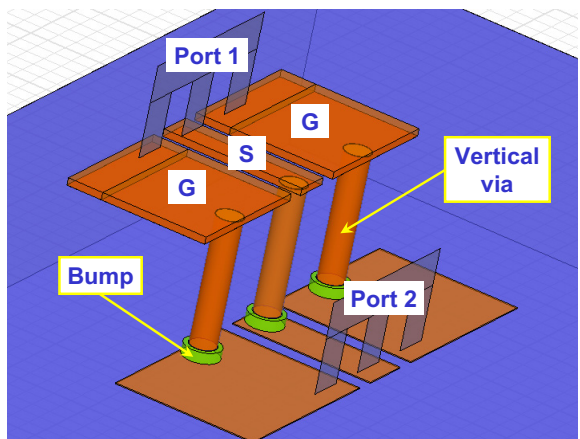

Fig. 11: Schematic view of a GSG through-vertical vias connected at both the top and bottom ends to short CPW's.

The intrinsic RF-MEMS varactor together with the lumped elements network is firstly validated against experimental data without including the electromagnetic influence of the package. To this purpose the Smith chart for the simulated and measured transmission parameter (S21) is shown in Figure 12 for a 0 V applied bias (varactor up-state). The good superposition of the two curves proves both the accuracy of the MEMS compact models and the effectiveness of the extracted parasitics network up to 10 GHz.

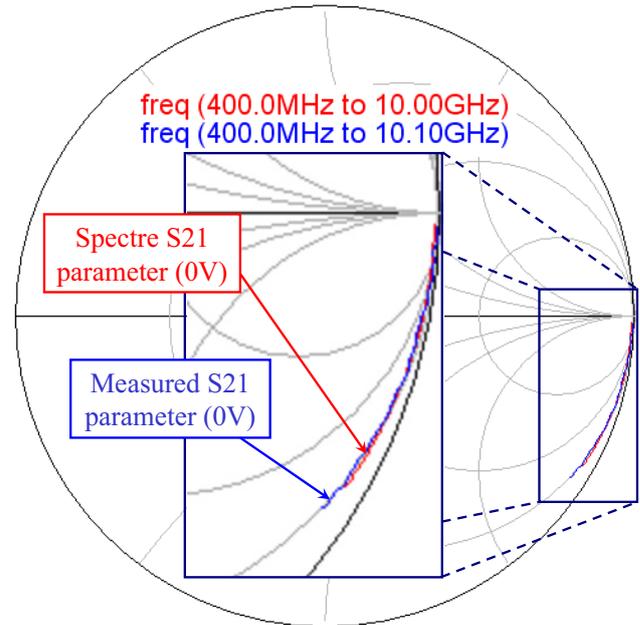

Fig. 12: Smith chart for the simulated (Spectre[©]) and experimental S21 parameter referred to the intrinsic RF-MEMS varactor surrounded by the extracted parasitics network.

The S-parameters of the uncapped MEMS varactor are compared with those obtained by the complete network of Figure 10. Spectre[©] simulation results are shown in Figures 13 (S11) and 14 (S21). Inclusion of vertical vias (see Figure 11) does not show a large influence on the S-parameters. For instance, the offset between the capped and uncapped S11 curves is about 3.6 dB at 6 GHz while, concerning the S21, it is 0.3 dB at 8 GHz. In this example only the effect of vertical vias plus short CPW's is considered, while the influence of the package-to-device wafers vicinity is not accounted for. In a real packaged structure the latter might have a not negligible role. However, we believe the presented approach can be easily generalized to achieve an accurate description also of complete devices.

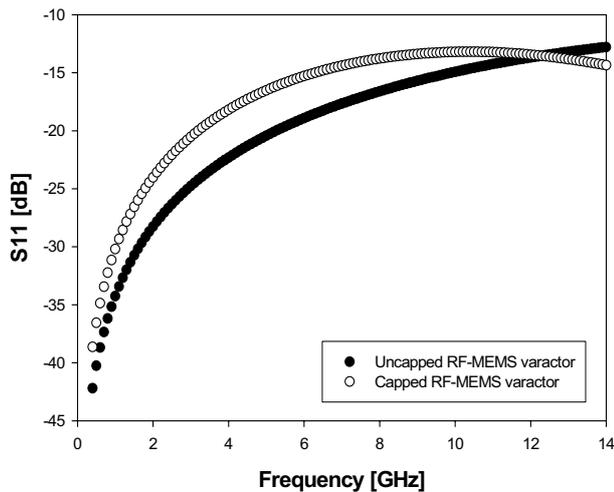

Fig. 13: Simulated reflection parameter (S11) of the capped vs. uncapped RF-MEMS varactor at 0 V bias.

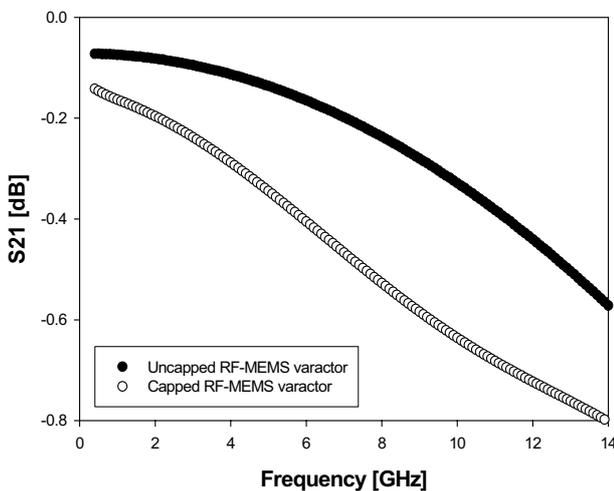

Fig. 14: Simulated transmission parameter (S21) of the capped vs. uncapped RF-MEMS varactor at 0 V bias.

### 5. CONCLUSION AND FURTHER WORK

In this work we presented our approach for characterizing and optimizing the RF performances of packaged RF-MEMS devices. Ansoft HFSS[TM] electromagnetic simulator is first validated against experimental data and then employed to optimize the technology degrees of freedom (DoF's) in the package fabrication through a parameterized model of a test capped CPW. The optimization led to the reduction of parasitic effects introduced by the package on the electromagnetic behaviour of capped devices. Subsequently, an RF-MEMS varactor was simulated in Spectre[©] by means of a MEMS compact model library in VerilogA[©] language. A lumped elements network surrounding the intrinsic varactor accounting for the parasitics was extracted from experimental data. Moreover, S-parameters of the capping through-vertical vias extracted from HFSS[TM] simulations were linked to the Spectre[©] schematic as S-parameters blocks. The discussed example is a proof of concept showing that it is possible to include all the significant effects influencing the RF behaviour of capped RF-MEMS devices within Spectre[©].